\def\be{\begin{equation}}
\def\ee{\end{equation}}
\def\ber{\begin{eqnarray}}
\def\eer{\end{eqnarray}}
\def\bers{\begin{eqnarray*}}
\def\eers{\end{eqnarray*}}

\def\JPC{J. Phys. C}
\def\JPCM{J. Phys.: Condens. Matter}
\def\PR{{ Phys. Rev.}\ }

\def\JPC{{ J. Phys. C: Solid State Phys}\ }
\def\JPCM{{ J. Phys.: Condens. Matter}\  }

\documentclass[aps,prb,twocolumn,showpacs,groupedaddressamsmath,amssymb]{revtex4}
\usepackage{dcolumn}   
\usepackage{amsmath}
\usepackage{graphicx}    
\usepackage{subfigure}
\usepackage{color}
\newcommand{\comment}[1]{}
\usepackage{ifthen}
\newboolean{includefigs}
\setboolean{includefigs}{true}      
\newboolean{includetext}
\setboolean{includetext}{true}     
\newcommand{\condcomment}[2]{\ifthenelse{#1}{#2}{}}
\begin{document}
\title[Green's function multiple-scattering theory with a truncated basis set: an augmented-KKR]{Green's function multiple-scattering theory with a truncated basis set:  An \emph{Augmented}-KKR formalism}
\author{Aftab Alam,$^{1\dagger}$ Suffian N. Khan,$^{2}$ Andrei Smirnov,$^{2}$ D.M. Nicholson,$^{3}$ and Duane D. Johnson,$^{2,4}$}
\email{aftab@phy.iitb.ac.in; ddj@ameslab.gov}
\affiliation{$^{1}$Department of Physics, Indian Institute of Technology, Bombay, Powai, Mumbai 400 076, India}
\affiliation{$^{2}$Division of Materials Science \& Engineering, Ames Laboratory, Ames, Iowa 50011 USA;}
\affiliation{$^{3}$Oak Ridge National Laboratory, Oak Ridge, Tennessee 37831 USA;}
\affiliation{$^{3}$Department of Materials Science \& Engineering, Iowa State University, Ames, Iowa 50011 USA.}

\begin{abstract}
Korringa-Kohn-Rostoker (KKR) Green's function, multiple-scattering theory is an efficient site-centered, electronic-structure technique for addressing an assembly of $N$ scatterers. Wave-functions are expanded in a spherical-wave basis on each scattering center and indexed up to a maximum orbital and azimuthal number $L_{max}=(l,m)_{max}$, while scattering matrices, which determine spectral properties, are truncated at $L_{tr}=(l,m)_{tr}$ where phase shifts $\delta_{l>l_{tr}}$ are negligible.
Historically, $L_{max}$ is set equal to $L_{tr}$, which is correct for large enough $L_{max}$ but not computationally expedient; a better procedure retains higher-order (free-electron and single-site) contributions for $L_{max}>L_{tr}$ with $\delta_{l>l_{tr}}$ set to zero [Zhang and Butler, Phys. Rev. B {\bf 46}, 7433]. 
We present a numerically efficient and accurate \emph{augmented}-KKR Green's function formalism that solves the KKR equations by exact matrix inversion [$\mathcal{R}^3$ process with rank $N(l_{tr}+1)^2$] and includes higher-$L$ contributions via linear algebra [$\mathcal{R}^2$ process with rank $N(l_{max}+1)^2$].
Augmented-KKR approach yields properly normalized wave-functions, numerically cheaper basis-set convergence, and a total charge density and electron count that agrees with Lloyd's formula.  
We apply our formalism to fcc Cu, bcc Fe and L$1_0$ CoPt, and present the numerical results for accuracy and for the convergence of the total energies, Fermi energies, and magnetic moments versus $L_{max}$ for a given $L_{tr}$.
\end{abstract} 
\date{\today}
\maketitle
\section{Introduction}
{\par}Multiple-scattering theory, as formulated by Korringa,\cite{KKR1} Kohn and Rostoker\cite{KKR2} (KKR), continues to be a powerful and efficient method to study the electronic structure of solids.\cite{Ebert.RPP2011} KKR theory is Rayleigh-Ritz variational, like related Muffin-tin Orbital (MTO) and Augmented Plane Wave (APW) methods. KKR Green's function (GF) techniques have facilitated numerous successful applications to spectral and energy related properties, such as surfaces,\cite{Ebert.RPP2011} alloys,\cite{KKR-CPA1,KKR-CPA2,KKR-CPA3,KKR-CPA4} interfaces,\cite{LKKR,KKR-CPA5} quantum criticality,\cite{KKR-CPA6} and transport.\cite{Transport1} Due to its inherent multiple-scattering nature, KKR-GF are used extensively to predict and analyze experimental results\cite{expt1} involving low-energy electron diffraction (LEED),\cite{expt2,expt3} photoemission,\cite{ARPES1,ARPES2,ARPES3} neutron and x-ray scattering.\cite{Scattering1,Scattering2,Scattering3}

{\par} A key parameter controlling KKR convergence is the maximum orbital and azimuthal number $L_{max}=(l,m)_{max}$ of the truncated spherical-wave basis on each scattering center. Historically, at a wave-vector $\bf{k}$ and energy $E$, $L_{max}$ was also chosen to control truncation of the single-site scattering $t_{LL'}(E)$ matrices and KKR scattering-path operator $\tau_{LL'}({\bf k};E)$ that dictates spectral properties of the system.
However, $\tau_{LL'}({\bf k};E)$, i.e., the Green's function ($G$), could be truncated at $L_{tr}<L_{max}$, where phase shifts $\delta_{l>l_{tr}}$ are negligible (set to zero), giving smaller matrices to invert if we could directly include the contribution of higher L's ( $L_{max} > L_{tr}$ ) via single-site and free-electron part of $G$.

{\par} For $L_{tr}$ equal to $L_{max}$, researchers find \emph{apparent} convergence in closed-packed systems using $l_{max}\sim3$. Yet, a key source of error is due to normalization of wave-functions ($\Psi$), affecting the charge density $\rho({\bf r})$ and density of states (DOS) $n(E)$ calculated from the Green's functions. As such, if $\Psi$ is not correctly normalized, the integrated DOS from the $L_{tr}$-truncated basis does not exactly reproduce total number of electrons in the system, and the Fermi energy E$_{\text{F}}$ is slightly incorrect. 
Also, $L$-truncation introduces error in the dipole matrix elements, which couple $l$ and $l\pm1$ states, needed for transport, electron-phonon, and atomic forces calculations.

{\par} So, a balance is struck between convergence of KKR-GF properties versus $l_{max}$ and numerical efficiency for inverting KKR matrices with rank $N(l_{max}+1)^2$. Butler\cite{Butler90} investigated the accuracy and convergence of multiple-scattering theory versus $l$ for two muffin-tin scatterers in a two-center expansion, showing solution can be made arbitrarily accurate at some numerically costly $l_{max}\rightarrow 60$. Zhang and Butler\cite{Zhang92} established a more proper procedure:  Solve the secular equation to $L_{tr}$ and retain $L_{max}>L_{tr}$ contributions with $\delta_{l>l_{tr}}$ set to zero -- yielding continuous and correctly normalized wave-functions and an electron count from the DOS that agreed with that from Lloyd's formula. This formalism was derived in real space, but never implemented for realistic materials. No  equivalent KKR-GF in reciprocal space was derived or tested. 

{\par} We present an augmented-KKR GF formulation in which matrices of rank ($\mathcal{R}$) $N(l_{tr}+1)^2$ are solved by direct inversion [$\mathcal{R}^3$ process] and contributions above $L_{tr}$ are included by a closed-form ($\delta_l\rightarrow 0$) linear algebra [$\mathcal{R}^2$ process] augmentation of the matrices to rank $N(l_{max}+1)^2$. Augmented-KKR yields normalized wave-functions, numerically fast basis-set convergence, and an electron count that agrees with Lloyd's formula. We tested convergence of total energy, E$_{\text{F}}$ (integrated DOS) and magnetic moments in three systems: fcc Cu, magnetic bcc Fe, and magnetic L$1_0$ CoPt. 

\section{FORMALISM}
In KKR-GF theory, the site-diagonal Green's function at a specific energy $E$ is given by
\begin{eqnarray}\label{eq1}
G({\bf r,r'},E)=\sum_{LL'} && [ Z^{n}_{L}({\bf r},E) \tau^{nn}_{LL'}(E) Z^{n}_{L'}({\bf r'},E) \\
&& - Z^{n}_{L}({\bf r},E) J^{n}_{L'}({\bf r'},E) \delta_{LL'}]  \nonumber ,
\end{eqnarray}
where $L$=$(l,m)$ for the site-centered, spherical-harmonic basis set. 
The tensor $\tau^{nn'}_{LL'}(E)$ is the scattering-path operator\cite{Gyorffy1971} describing the propagation pathway of electrons in an array of scattering centers. 
$Z^{n}_{L}({\bf r},E)$ and $J^{n}_{L}({\bf r},E)$ are, respectively, the regular and irregular solutions of the Schr$\ddot{o}$dinger equation in the $n$-th Wigner-Seitz cell. $Z^{n}_{L}({\bf r},E)$ has the form
\begin{eqnarray}
Z^{n}_{L}({\bf r},E)  = \kappa \sum_{L'} \phi^{n}_{L'}({\bf r},E) [S^{n}_{L'L}(E)]^{-1}  ,
\label{eq2}
\end{eqnarray}
where $\kappa=\sqrt{E-v_0}$. Here, $v_0$ is an arbitrary reference energy for an exact theory, but, for approximate cases, such as muffin-tin (MT) or atomic sphere approximations (ASA), it can be chosen variationally to match the trace of eigenvalues of the exact theory.\cite{Alam2009}
$\phi^{n}_{L'}({\bf r},E)$ is the wave-functions solution with potential $v_n({\bf r})$, i.e.,
\begin{eqnarray}
\left[ -\nabla^2 + v_{n}({\bf r}) \right] \phi^{n}_{L}({\bf r},E) = E \phi^{n}_{L}({\bf r},E)  .
\label{eq3}
\end{eqnarray}
The potential vanishes outside the convex cell, so $\phi^{n}_{L}$ joins smoothly to a combination of spherical Bessel $j_l(\kappa r)$ and Neumann $n_l(\kappa r)$ functions beyond the circumscribing sphere (CS) radii around the cell ($r>R_{CS}$), i.e.,
\begin{eqnarray}
\phi^{n}_{L}({\bf r},E) = \sum_{L'} \left[ n_{l'}(\kappa r) S^{n}_{L'L}(E) - 
j_{l'}(\kappa r) C^{n}_{L'L}(E) \right] Y_{L'}(\widehat{r})\nonumber\\
\label{eq4}
\end{eqnarray} 
The sine {\bf S} and cosine {\bf C} matrices are calculated by matching the continuity of the logarithmic derivative of $\phi^{n}_{L}$ across the cell boundary. Notably, $J^{n}_{L}({\bf r},E)$ has the asymptotic limit 
\begin{eqnarray}
J^{n}_{L}({\bf r},E) \rightarrow j_{l}(\kappa r) Y_{L}(\widehat{r})\ \ \ \ ; \ \ \ \ r>R_{CS}
\label{eq5}
\end{eqnarray}

\subsection{ KKR-GF formalism } While constructing $\tau^{ij}_{LL'}(E)$, the propagation of electrons from one scattering center $i$ to another $j$ is defined by the free-electron Green's function ${\bf g}^{ij}_{LL'}(E)$ (in a spherical-harmonic basis), or in a solid the KKR structure constant matrix ${\bf g}^{nn'}_{LL'}({\bf k};E)$ with basis sites on $n,n'$ sublattices. In a solid, with periodic boundary conditions invoked,  $\tau^{nn'}_{LL'}({\bf k};E)$ is given in finite matrix form as
\begin{eqnarray}
{\bf \tau=(1-tg)^{-1} t = t + tgt + tg\tau gt } ,
\label{eq6}
\end{eqnarray}
where ${\bf t}$ is the single-site scattering matrix, which in a cell $n$ is generally given by
\begin{eqnarray}
{\bf t}^{n}=-{\kappa}^{-1}({\bf C}^n - i {\bf S}^n)^{-1}{\bf S}^n  .
\label{eq7}
\end{eqnarray}
For a spherically symmetric scatterer\cite{Messiah} (used here), the single-site $t$-matrices simplify as
\begin{eqnarray}
t_{LL'}(E) \rightarrow t_{l}(E)\delta_{LL'} = -\frac{1}{\kappa} sin~\delta_l(E) e^{i\delta_l(E)} .
\label{eq7p}
\end{eqnarray}
For MT or ASA scattering centers, the KKR phase-shifts are determined by matching the free-electron solution on the sphere boundary.

{\par}Generally, using Eq.~\ref{eq6}, the full GF (\ref{eq1}) can be rewritten in terms of single-site and multiple-scattering pieces (in matrix form), i.e.,
\begin{eqnarray}\label{eq8}
{\bf G}({\bf r,r'},E) &=&  {\bf (ZtZ - ZJ)} + {\bf Z(\tau - t)Z} \nonumber\\
&=& {\bf (ZtZ-ZJ) + Zt (g) tZ + Zt (g\tau g) tZ} . \nonumber \\
\end{eqnarray}
Each quantity above is a super matrix in a space of angular momentum [rank $(l+1)^2$] and of unit cell size $N$, giving a total rank of $\mathcal{R}=N(l+1)^2$. 

{\par} The three major computational expenditures in KKR-GF theory are calculations of (1) structure constants {\bf g}; (2) wave-functions {\bf Z} and {\bf J};  and, most costly, (3) {\bf $\tau$} from Eq.~\ref{eq6}, which requires an $\mathcal{R}^3$ operation for the inversion. Now, with $N$ fixed, $L$ is usually truncated in numerical calculations to a small, but necessary value (e.g., $L=3$) above which the phase-shifts $\delta_l$ are assumed to be zero, but which is not an $L$ where the higher-order terms can necessarily be ignored -- an error.

\subsection{Augmented-KKR-GF} While free-electron contributions remain at all $L$'s, the phase shift $\delta_l$ for a spherical scatterer decays rapidly (at standard temperatures and pressures) with increasing value of $L$. Thus, while the first line of Eq.~\ref{eq8} is convenient numerically  (e.g., for pole cancellation and contour integration, and finite-temperature Matsubara sums\cite{Johnson1984,Johnson1985,Pinski1985}), the second line provides a simple means to account for KKR multiple-scattering solutions exactly the same way as in the conventional KKR-GF theory up to $L_{tr}$ and then augment with single-site and free-electron contributions from $L_{tr} < L \le L_{max}$, while maintaining symmetry and relative accuracy.  

{\par} So, in augmented-KKR, we analytically evaluate Eq.~\ref{eq8} to include $L > L_{tr}$ (in the limit $\delta_l \rightarrow 0$) terms via linear algebra, rather than full matrix inversion.  First, $g_{LL'}$ is calculated for $L \le L_{max}$ to where augmentation is desired. Second, for $L \le L_{tr}$, the terms in Eq.~\ref{eq8} are evaluated as usual, while, for $L > L_{tr}$ in $\delta_l \rightarrow 0$ limit, the first two terms can be analytically simplified using, 
\begin{eqnarray}
{\bf Z t} &\xrightarrow[\delta_{l} \rightarrow 0]{}&   + {\bf j} (\kappa r)\nonumber\\
{\bf (Z t Z - Z J)} &\xrightarrow[\delta_{l} \rightarrow 0]{}&  -\kappa\ {\bf j}(\kappa r) \left[ i\ {\bf j}(\kappa r)  -  {\bf n}(\kappa r) \right].
\label{eq9}
\end{eqnarray}
Equation~\ref{eq9} is derived (see Appendix) rigorously using expressions for spherical potentials, which vanish outside of spheres inscribed within each cell. They do not hold for full-cell potentials, where non-diagonal $L,L'$ terms can contribute generally, but can be derived.

{\par} Lastly, the most crucial step is evaluating the last term in Eq. \ref{eq8}. Positing negligible scattering for large $L$'s, the last term is calculated in three steps:
\begin{enumerate}
\item Calculate  $\tau_{L_1L_2}=[{\bf (1 - tg)^{-1}t}]_{L_1L_2}$ for $L_i \le L_{tr}$ by exact inversion. 
\item With $g_{LL'}$ ($\forall\ L,L'=L_{max}> L_{tr}$), calculate ${\bf g \tau g}$ using;
\begin{eqnarray} \label{eq10}
({\bf g \tau g})_{LL'} &=& \sum_{L_1}^{L_{tr}} \sum_{L_2}^{L_{tr}} g_{LL_1} \tau_{L_1L_2} g_{L_2L'} 
\end{eqnarray}
\item Having ${\bf g \tau g}$, multiply $({\bf Zt})_{LL'}$ from both sides to get G$({\bf r,r'},E)$ for all $L=L_{max}$.
\end{enumerate}
With this, one needs to perform inversion ($\mathcal{R}^3$ operation) only for matrices up to $L_{tr}$, and the higher-$L$ contributions are included by matrix multiplication (${\bf g \tau g}$), which is computationally much faster ($\mathcal{R}^2$ operation).

{\par} In an all-electron, {\it ab initio} calculation for real systems, $l$ truncation enters at several places and collectively affects, e.g., the cell DOS, and charge and magnetization densities. In turn, the Fermi energy E$_{\text{F}}$ and magnetization $\text{M}$, defined from the sum rules, 
\begin{eqnarray}
\text{N}({\text{E}}_{\text{F}}) &=& \int_{E_{bot}}^{E_{\text{F}}} \left[ n_{\uparrow}(E') + n_{\downarrow}(E')\right] dE' = Z_{val}\nonumber\\
\text{M} &=& \int_{{E}_{bot}}^{E_{\text{F}}} \left[ n_{\uparrow}(E') - n_{\downarrow}(E')\right] dE'
\label{eq11}
\end{eqnarray}
are affected, as is the total energy. Here, $E_{bot}$ designates the bottom of the valence band, $n_{\uparrow} (n_{\downarrow})$ is the spin majority (minority) DOS, and $Z_{val}$ is the average number of valence electrons. As will be shown, $l$ truncation plays a significant role in correctly evaluating the E$_{\text{F}}$ and $\text{M}$. 

\section{Computational Details}
{\par }An all-electron, density functional theory (DFT) KKR-GF code\cite{MECCA} is used to perform the calculations, as previously done.\cite{KKR-CPA1,KKR-CPA4,KKR-CPA5} For the present results, the von-Barth--Hedin\cite{von-Barth72} local spin-density approximation (LSDA), as parameterized by Moruzzi, Janak and Williams,\cite{MJW78} was used. 
Each site-dependent Voronoi polyhedra were represented within an ASA sphere,\cite{Phar84} with multi-component cases handled by an optimal basis,\cite{Alam2009} where ASA spheres are adjust by saddle-points in the electronic density.  Complex energy contour integration  with $24$ energy points are used to integrate the Greens function. 
Monkhorst and Pack\cite{Monk76} special {\bf k}-point method is used for Brillouin zone integration.

{\par} Following the above theory, in distinction to conventional KKR where $L$ is truncated where $\delta_L(E)\approx0$ (rather than where free-electron contributions are small, which is $E$, $L$, and temperature dependent), two distinct  $L_{tr}$ and $L_{max}$ indices are used. All the calculations up to $L_{tr}$ are performed in the standard way, i.e., for each energy $E$, we evaluate (${\bf Z}, {\bf J}$), ${\bf t}$ and ${\bf g}$ and get ${\bf \tau}$ by inversion. 
For augmented-KKR, beyond $L_{tr}$, we calculate the truncated ${\bf \tau}$ for $L,L' \le L_{tr}$, and use the full $g_{LL'}$-matrix to augment ${\bf g}+{\bf g}{\bf \tau} {\bf g}$ Eq. \eqref{eq10} up to $L_{max}$, which can be chosen manually or to be below a specified tolerance for GF error.
Recall from Eq. \ref{eq8}, we know the analytic form of ${\bf Z t Z - Z J}$ and ${\bf Zt}$ for $L>L_{tr}$, so the only effort in evaluating the matrix elements of ${\bf g}+{\bf g}{\bf \tau} {\bf g}$ for $L_{tr} < L \le L_{max}$.

\section{Results and Discussion}
{\par} We apply the {\it ab-initio} augmented-KKR formalism to fcc Cu, bcc Fe, and L$1_0$ CoPt. The first two systems provide a stringent test of our formalism for a simple non-magnetic and magnetic system, respectively. The third  illustrates application to a multi-sublattice, magnetic example.

\subsection{Convergence of Total Energy and Fermi Energy}
\begin{figure}[t]
\centering
\includegraphics[width=8cm]{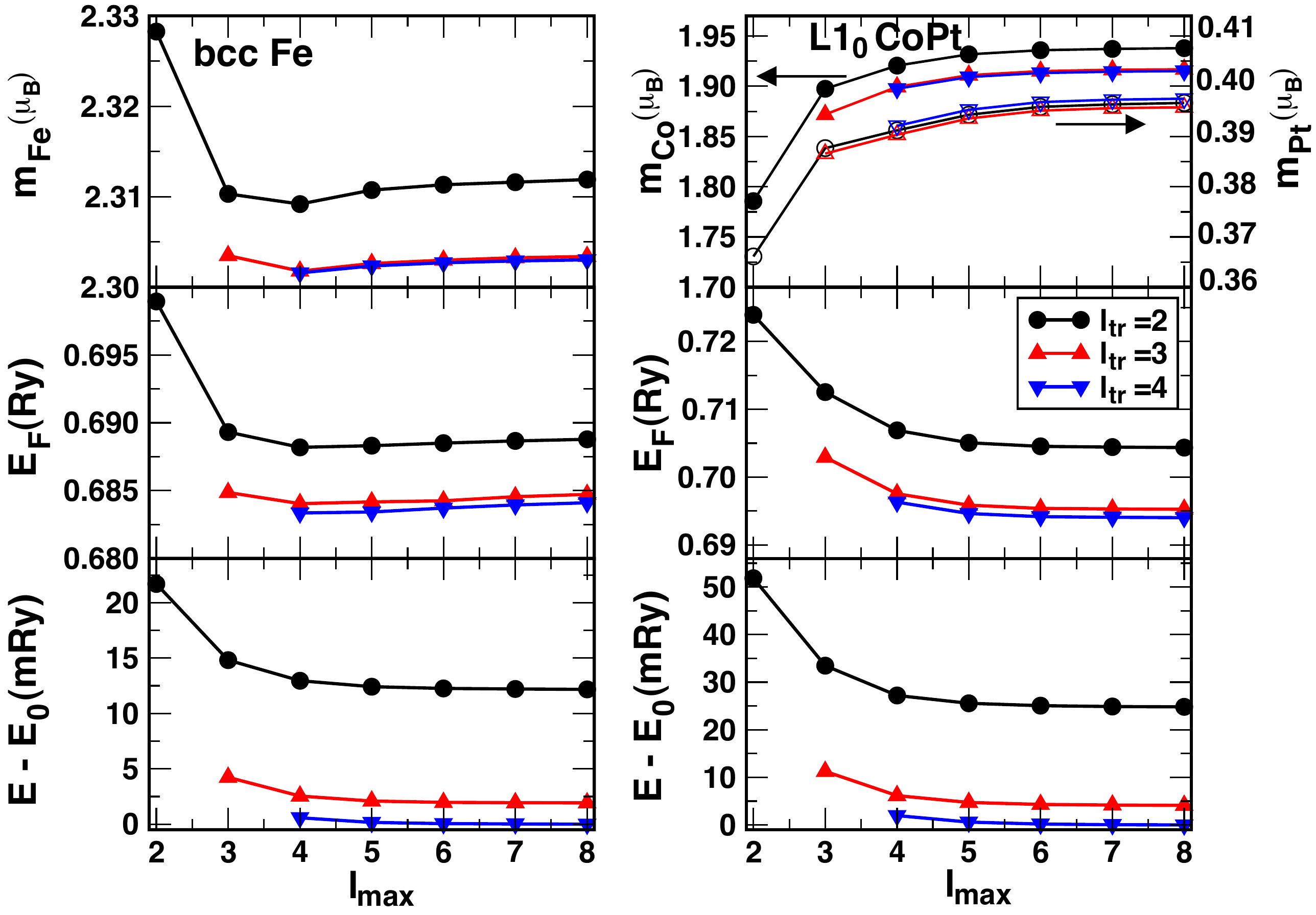}
\caption {(Color online) (Left) Convergence properties of total energy E (bottom) and Fermi 
level E$_\text{F}$ (top) vs. $l_{max}$ using different $l_{tr}$ for $1$-atom/cell fcc Cu. 
(Right) Same as left, but for a $1$ Cu-site plus $1$ octahedral-hole-site per cell. E$_0$ is defined by the $l_{tr}=4,l_{max}=8$ result. E$_{\text{F}}$ is different on the left because of the Madelung potential inherent on the right. }
\label{fig2}
\end{figure}

{\par} Figure \ref{fig2} shows the convergence of total energy (bottom) and Fermi energy (top) versus the augmentation $l_{max}$, with $l_{tr}=2,3,4$ for fcc Cu in our KKR-ASA code.\cite{MECCA} The left (right) panel indicate the results for a basis of one Cu atom (atom + octahedral hole). The right panel shows improvement both in augmentation and basis set, as the hole-site makes the Voronoi polyhedra for each scattering site (atom and hole) more spherical, and the atom is better represented by an ASA sphere (and reduces the ASA overlap error), while the interstitial volume is greatly reduced.

{\par} The energies are not converged with $l_{tr}=2$ and large $l_{max}$ compared to converged values. By $l_{max}=5$ the total energy (E$_\text{F}$) reaches an asymptotic value, and $l_{tr}=2$ value is higher by $9\ m$Ry ($4\ m$Ry) relative to those calculated with $l_{tr}=4$. Beyond $l_{max}=5$, the error in total energy  (E$_\text{F}$) is less than $0.05\ m$Ry  ($0.01\ m$Ry), i.e., the order of $1 \times 10^{-5} Ry$.  For $1$-atom Cu basis (left panel), the total energy (E$_\text{F}$) with $l_{tr}=4$ converged to within $1\ m$Ry ($0.1\ m$Ry) compared to that with $l_{tr}=3$. With an octahedral hole added to the basis  (right panel) the results are exactly the same for $l_{tr}=3$ or $4$. 
This is due to the use of a better basis set. As such, a faster convergence in the $l$-space can be achieved by an improved basis set, but with a concomitant increase in $\mathcal{R}^3$ process. 

{\par} Moghadam {\it et al.}\cite{Moghadam2001} carried out a test of $l$ convergence for fcc Cu in a real-space KKR, linear-scaling multiple-scattering (LSMS) method. The difference between the total energy (E$_\text{F}$) for $l_{max}=3$ and $l_{max}=8$ is $7\ m$Ry ($6\ m$Ry).
For the {\bf k}-space-based, augment-KKR the differences are $6.5\ m$Ry ($8\ m$Ry) for 1-Cu-site basis, and
$5\ m$Ry ($7\ m$Ry) for a basis with a Cu plus an octahedral hole. The present method, however, is  computationally faster due to the augmentation used to evaluate the contribution of higher $l$'s.

\begin{figure}[t]
\centering
\includegraphics[width=8.5cm]{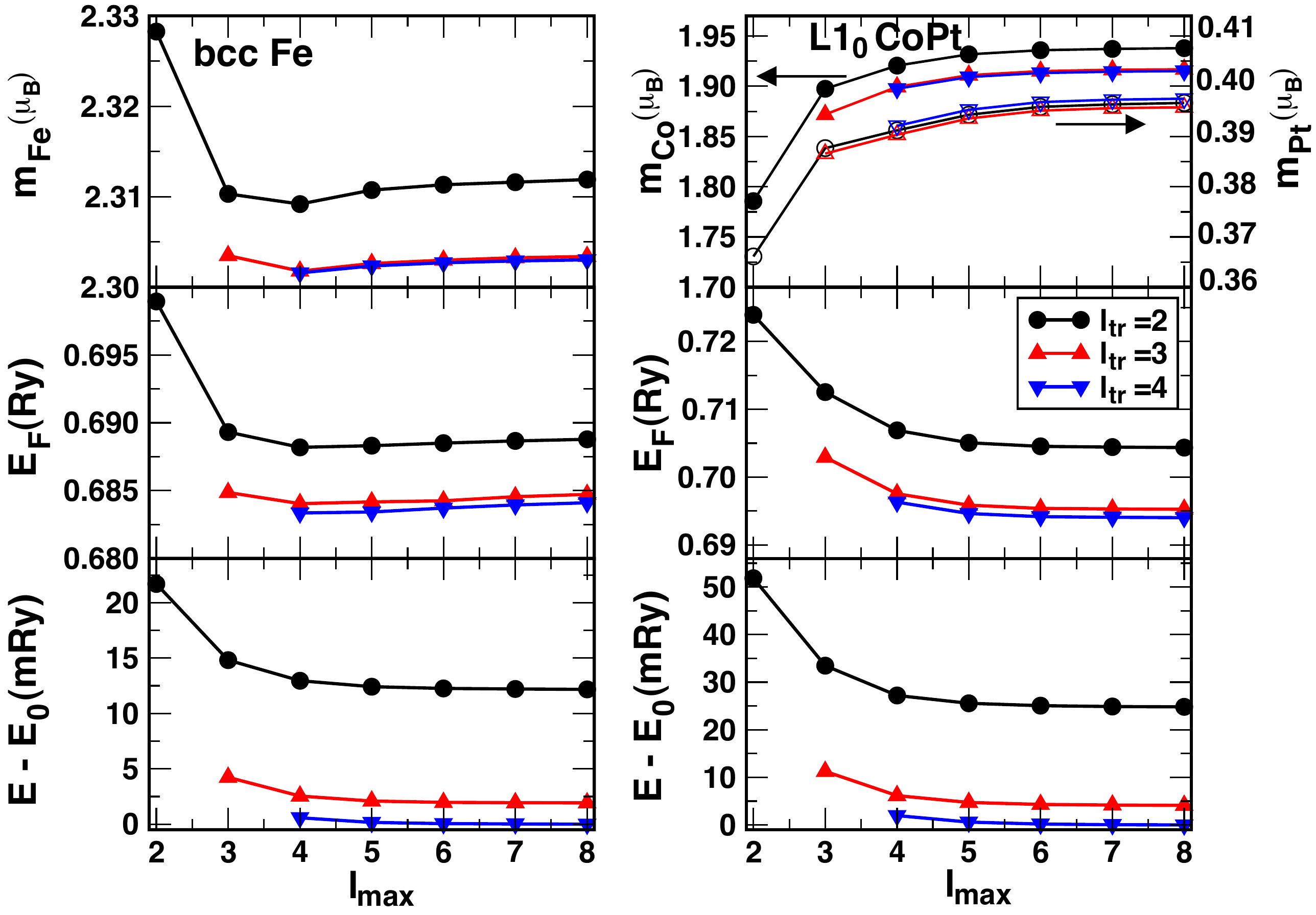}
\caption {(Color online) (Left) Convergence of total energy E (bottom), E$_\text{F}$ (middle) and magnetic moments (top) vs. $l_{max}$ at different $l_{tr}$ for a $1$-atom bcc Fe, and (right) for a $2$-atom L$1_0$ CoPt. E$_{0}$ is $L_{max}=8$ reference value.}
\label{fig3}
\end{figure}

{\par} Figure~\ref{fig3} shows the convergence of total energy (bottom), E$_\text{F}$ (middle) and magnetic moments (top) for magnetic bcc Fe (left panel) and L$1_0$ CoPt (right panel). As in the case of fcc Cu, total energy and E$_\text{F}$  converges by $l_{max}=5$ with $l_{tr}=4$.  The converged moment of bcc Fe is $2.30$ $\mu_{B}$, which compares well with experimental value\cite{expt_bcc} of $2.2$ $\mu_{B}$. The Co and Pt moment in L$1_0$ CoPt converges relatively slower compared to that of Fe. This is due to a slight $c/a$ distortion in the L$1_0$ structure ($c/a=0.984$). Calculated moments for Co and Pt are $1.91$ and $0.396$ $\mu_B$, respectively, compared to $1.76$ and $0.35$ $\mu_B$ from experiment at finite temperature.\cite{Grange2000}

\begin{figure}[b]
\centering
\includegraphics[width=6.5cm]{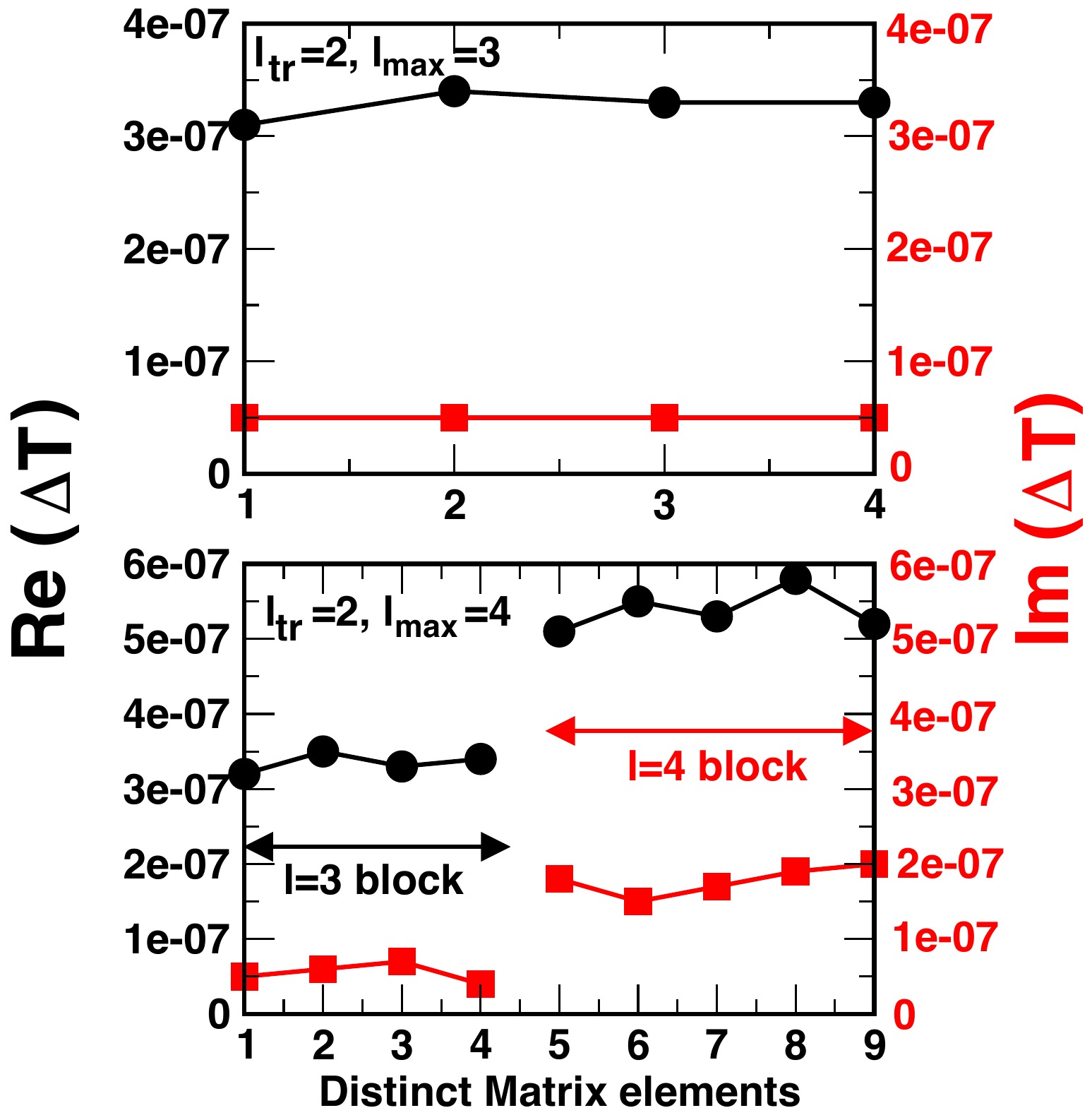}
\caption {(Color online) Absolute difference in distinct elements of T$_{LL'}=(g+g\tau g)_{LL'}$ calculated using full $\tau_{LL'}$ with $L,L' \le L_{max}$ and the augmented-KKR $\tau_{LL'}$ with $L,L' \le L_{tr}=2$ and $L_{max} =3$ (top) and $L_{max} =4$ (bottom) at $E=(-0.76,0.003)$~Ry. Similar accuracy is found along the entire semi-circular contour of integration. For spherically symmetric scatterers with $L_{max}=3$ ($4$), there are $4$ ($5$) distinct matrix elements in the $L=3$ ($4$) block of the T-matrix.   }
\label{fig1}
\end{figure}

{\par} Lastly, for comparison of matrix elements from full-KKR versus augmented-KKR, we first calculated ${\bf \tau}$ for $L,L' \le L_{max}$ by direct inversion, and then calculated ${\bf T}={\bf g}+{\bf g}{\bf \tau}{\bf g}$ matrix using both full and truncated $\tau$ matrices.   Figure \ref{fig1} shows the absolute error ($\Delta$T) in the matrix elements of the larger ($L> L_{tr}$) block of augmented T-matrix calculated using full $\tau$ and truncated $\tau$ for fcc Cu. With $l_{tr}=2$, the augmentation is compared for $l_{max}=3$ ($l_{max}=4$) in the top (bottom) panel. For fcc symmetry, there are four (five) distinct matrix elements in the $l=3$ ($l=4$) block of the T-matrix, which are labeled along the horizontal axis. Clearly, augmented-KKR well reproduces the higher $l$-block of T-matrices compared to full-KKR inversion (quite well below $1 \times 10^{-6}$), showing that the computationally faster augmentation has very good accuracy.

\subsection{Comparison to Lloyd's Formula}
{\par} The Lloyd's formula is the N-site generalization  of the Friedel (single-site) sum-rule, or optical theorem, for the electronic integrated DOS:
\begin{align} \label{eq:sumofshifts}
N(E)=N_\text{free}(E)+\frac{2}{\pi} \sum_\ell (2\ell+1)\delta_\ell(E)\,,
\end{align}
where $N_\text{free}(E)$ is the integrated DOS of the free electrons, known analytically in 3-dimensions, i.e., $E^{3/2}/(6\pi^2$). [For complex $E$, Eq.~(\ref{eq:sumofshifts}) is incorrect, but can be generalized via scattering matrices.]
The KKR Lloyd's formula for an ordered system can be written in spectral representation at any complex $E$ as\cite{Lloyd1967,Lloyd1972,Zeller2004} 
\begin{align}\label{eq:lloyd-kkr}
N(E) &=& N_\text{free}(E) + \frac{1}{\pi} \text{Im log}||{\bf \alpha}(E)|| \nonumber \\ 
     \,\,\,\,\,\, && -\frac{1}{N_k}\Sigma_{\bf k}\frac{1}{\pi}\text{Im log}||1-{\bf t}(E){\bf g}({\bf k},E)||
\end{align} 
for discrete samples in ${\bf k}$-space, with $\alpha(E)$ defined by the scattering solutions of $v(r)$ near the scattering centers. The determinant is performed over both basis-site and angular-momentum indices, and it is equivalent to an eigenvalue sum, albeit done by constant-E scan.  The formula is also related to Krein's theorem.\cite{Krein1953,Faulkner1977} Lloyd's formula is an amazing result being the closed-form expression for the integrated DOS at any $E$ (total electrons)! Moreover, variation of Lloyd's formula with respect to potential ${v}({\bf r})$ yields the density at any energy $\rho({\bf r};E)$ to second-order in changes in the self-consistent potential (e.g., higher-order L's), and, hence, in E$_\text{F}$.\cite{KKR-CPA1}

{\par} Because the KKR determinant passes through zero at every Bloch solution, it picks up a phase of $\pi$ at these locations giving the number of electrons up to $E$. Equation (\ref{eq:lloyd-kkr}) counts jumps in phases in the KKR determinant (``Im log'' operation). There is a practical implementation issue: \emph{At a given E} the phase is known to modulo $2\pi$ (or total electrons within a whole number), but trivially handled with a good E$_\text{F}$ estimate from real-space GF. Thus, Lloyd's formula gives an exact (no L-truncation) E$_\text{F}$ and electron count from a few values of $E$.

{\par} Importantly, here, the augmented-KKR is expected to yield a E$_{\text{F}}$ consistent with that obtained by Lloyd's formula.  Also, for thermodynamics of a system, an analytic expression for the free-energy functional can be directly derived from Lloyd's formula using a Gibbs relation,\cite{KKR-CPA1} which we use to calculate the total (free) energy. At finite temperature it directly yields Mermin's theorem, or Kohn-Sham theorem at zero Kelvin.\cite{KKR-CPA1} Hence, the spectral Lloyd's formula specifies the thermodynamics and correct Fermi surface at E$_\text{F}$.

\begin{figure}[t]
\centering
\includegraphics[width=8.7cm]{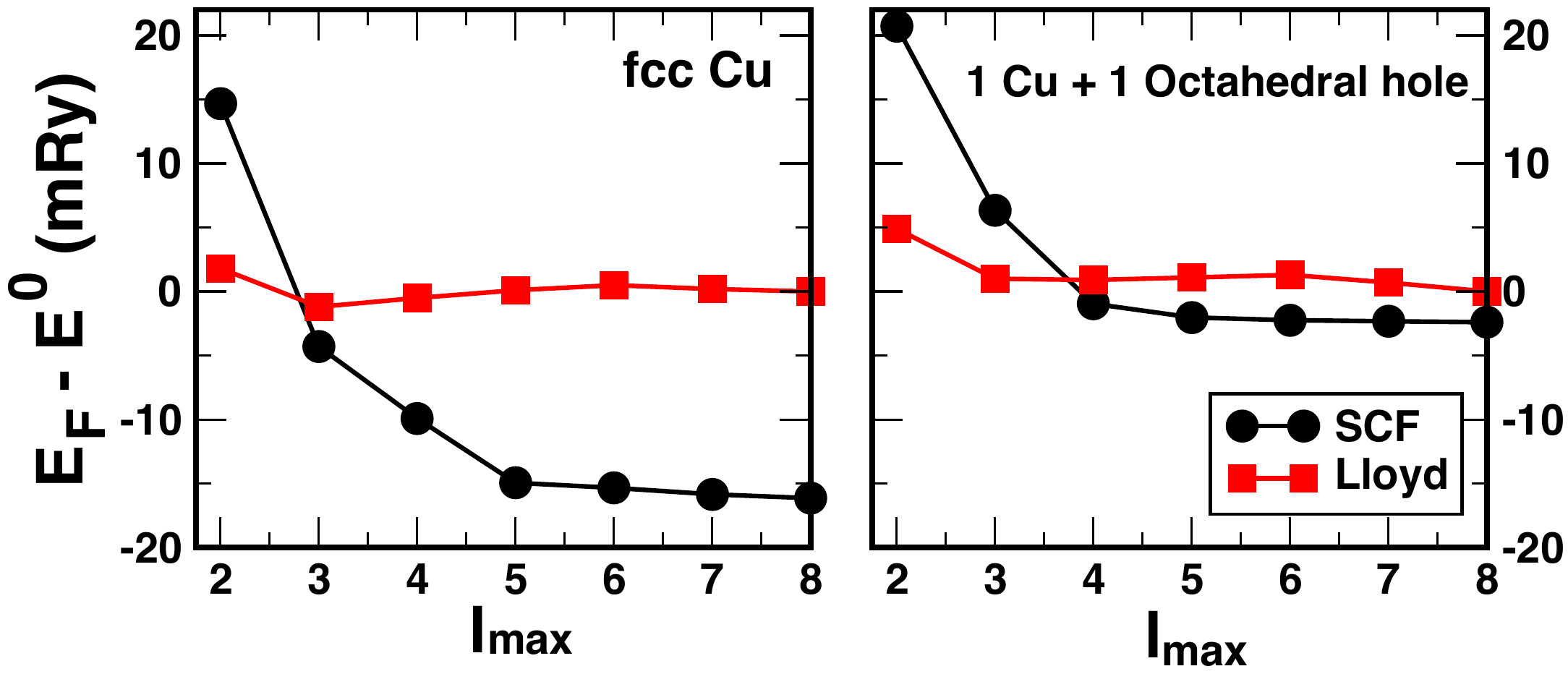}
\caption {(Color online) E$_\text{F}$ by real-space GF (circle) relative to spectral Lloyd's formula (square) for Cu (left) and Cu plus hole (right) for $L_{tr}=L_{max}$. E$^0$ is a reference energy.}
\label{fig4}
\end{figure}

{\par} To assess the augmented-KKR E$_\text{F}$ and electron count, we must compare three results: (1) Convergence of augmented-KKR spectral GF -- Lloyd's formula;  (2) Convergence of augmented-KKR real-space GF, as given by the trace of Eq.~\ref{eq8}; and (3) Convergence of items 1 and 2 with improved  basis. Notably, approach (2) is used in a typical self-consistent-field (scf) KKR-GF for computational expediency because {\bf G}({\bf r},{\bf r}';E) is always handy.  

{\par} To be clear, results from (1) and (2) should agree for an exact method. However, these values will differ if there is any approximation that is not handled equivalently in k-space and real-space -- as in the ASA.  For (1) above, {\bf G}({\bf r},{\bf r}';E) is evaluated within an ASA sphere and then Fourier transformed to obtain {\bf G}({\bf k},E) commensurate with the Brillouin zone of the full unit cell; hence, for ordered systems, it is a calculation of the volume enclosed by the Fermi surface, and corresponds to the count over a \emph{non-spherical} charge distribution. In contrast, for (2), the trace of {\bf G}({\bf r},{\bf r}';E) is evaluated in an ASA sphere, which does not account for the volume as done in (1), and it must suffer a modest error because only a spherical charge density is considered.  Therefore, (1) above should be correct, and a small error may appear from (2), which will decrease with, say, improving basis. 

{\par}For comparison, a wavefunction approach using the ASA solves the secular equation after Fourier transform by diagonalization to get the eigenvalues, and E$_\text{F}$ is then obtained by counting states within volume enclosed by the Fermi surface, then the remaining quantities are determined by referencing only k-space. As such, it is equivalent to the value from spectral Lloyd's formula, Eq.~\ref{eq:lloyd-kkr}.

{\par}From the scf-KKR , Figure \ref{fig4} compares E$_\text{F}$ versus $L_{max}$ (=$L_{tr}$) from real-space Green's function (circle) and from spectral (k-space) Lloyd's formula (square) for basis with one Cu (left) and for one Cu plus an octahedral-hole (right). These results agrees with those from augmented-KKR with $L_{tr}=3$ but evaluated for each $L_{max} \le 8$. As is apparent, the spectral Lloyd's results converge rapidly, whereas those from real-space converge slower and suffer a small error because only a spherical charge density is considered. As is obvious from Fig. \ref{fig4}, the discrepancy in E$_{\text{F}}$ obtained from the two methods reduces when an octahedral hole is inserted into the fcc cell (right panel) because the ASA then better represents the real-space volume; the k-space result is also improved because the non-spherical charge density is better represented. 

{\par} As will be discussed elsewhere (e.g., for applications to warm-dense matter), further improvements to the agreement between real-space and k-space results (not shown) are possible, which leads to much less than $1~m$Ry discrepancy. Example changes include: improve normalization of scattering functions ({\bf Z, J}), and include $N_\text{free}(E)$ analytically (infinite $L$ sum), as in Lloyd's formula, while simultaneously removing the $L$-truncated free-electron Green's function contributions from the KKR Green's function during the scf-cycle.   

\subsection{Convergence of Structural Parameters}
\begin{figure}[b]
\centering
\includegraphics[width=5.5cm]{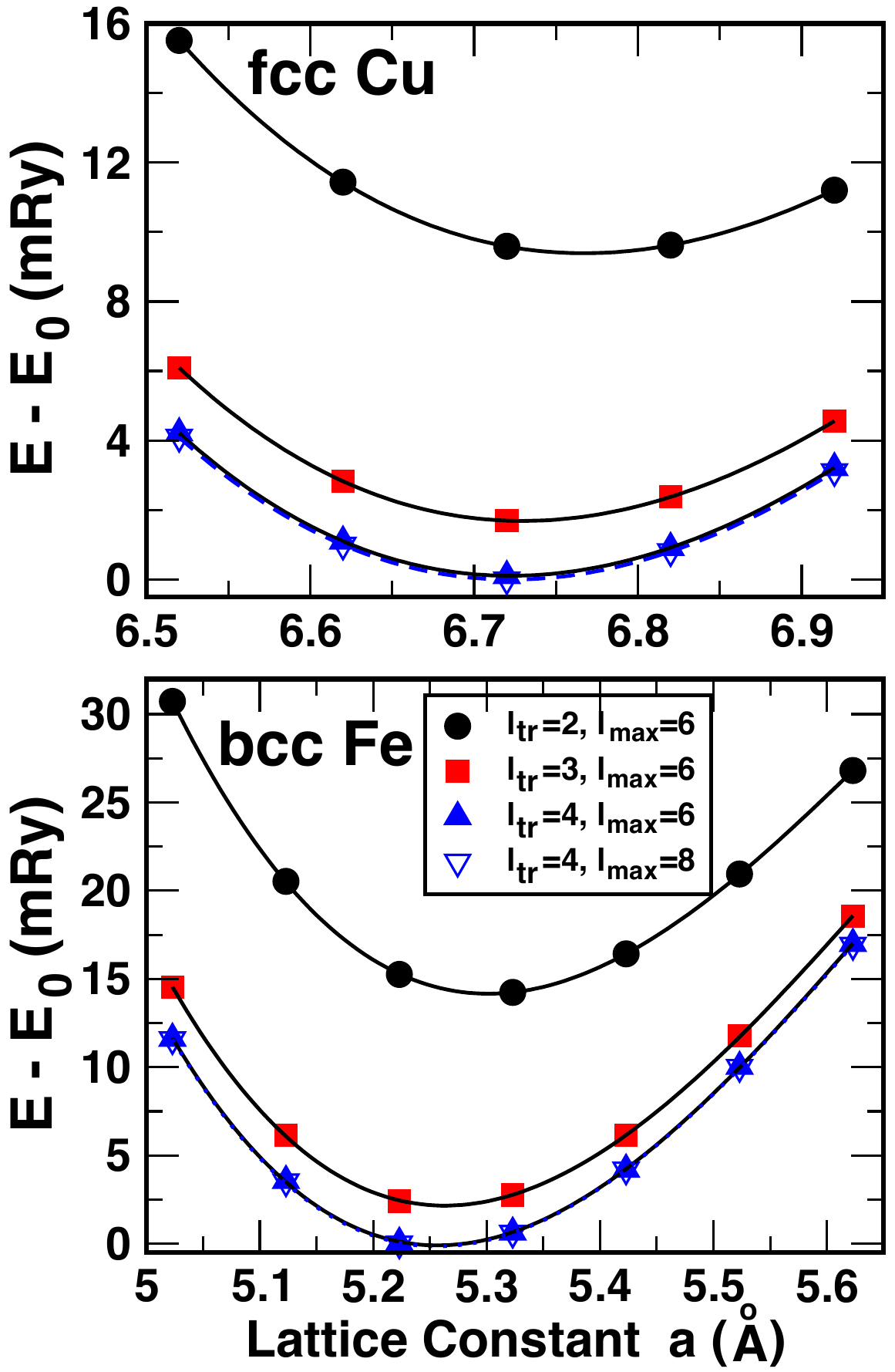}
\caption {(Color online) Total energy vs. $a$ for various sets of augmentation $(L_{tr},L_{max})$ for  Cu (top) and Fe (bottom).}
\label{fig5}
\end{figure}

{\par} Up to now we have investigated the convergence properties at a fixed lattice parameter. Another important thing to check is how convergence affect the accuracy of equilibrium (groundstate) lattice parameter. Figure \ref{fig5} shows the energy versus lattice constant ($a$) for fcc Cu (top) and bcc Fe (bottom) at different sets of $l$ truncation. Notably, for both the systems, the energy curve with $l_{tr}=4$ and $l_{max}=6$ is almost indistinguishable from that with $l_{max}=8$, indicating the convergence by $l_{max}=6$.
Already $l_{tr}=3$ finds a similar minima to that from larger $l_{tr}$; however, $l_{tr}=2$ is not a reliable choice for converged results. The calculated $0~K$ lattice constants for fcc Cu and bcc Fe are $6.72$ and $5.26$ {a.u.}, respectively, which compare well with room-temperature experimental values ($6.82$ and $5.42$~a.u., respectively), and previous LSDA results.

\subsection{Estimate of Numerical Savings}
In the current implementation the augmented-KKR requires $\left[ N^3(l_{tr}+1)^3 +  N^2(l_{max+1}+1)(l_{tr}+1) \right]$ operations compared to $\left[ N(l_{max}+1) \right]^3 $ operations in standard KKR. For reliable convergence in the present examples, we found $l_{tr}=3$ and $l_{max}=8$ to be sufficient for lattice constants and structural energy differences, in which case we require ($64 N^3 + 36 N^2$) operations as opposed to $278 N^3$ operations. Hence, about $3-4$ times less computational time is required for cells with 1-10 atoms. An estimate that holds the calculations done here.

\section{Conclusion}
{\par} Motivated by numerically efficient and physics, we have presented and successfully implemented an augment-KKR Green's function formalism that permits accurate handling of multiple-scattering by direct inversion of smaller $L_{tr}$-truncated basis (where phase shifts are not zero) and include higher $L>L_{tr}$ by linear algebra for necessary single-site and free-electron contributions. We applied the augmented-KKR formalism to three systems and showed very good accuracy and convergence properties, although a larger $L$-basis is needed over that generally assumed. 
To be mathematically consistent, the truncation of $L$-sum for the wavefunctions and the scattering matrices needs to be done in tandem with each other (see Eq. \ref{eq8}) due to a normalization factor occurring in both the single-site wavefunction and ${\bf t}$-matrices. By identifying this common normalization, one can analytically evaluate the $\delta_l \rightarrow 0$ limit to include higher $L$'s via simple linear algebra, instead of the exact inversion as required in conventional KKR, saving significant computational effort while improving accuracy.  Augmented-KKR can be extended to the coherent potential approximation (CPA) and dynamical cluster approximation (DCA) to handle disorder, as will be presented elsewhere.

\section{Acknowledgement}
AA acknowledges support from SEED Grant (sponsored project 13IRCCSG020) at IIT Bombay.
Work was also supported by the U.S. Department of Energy, Office of Science, Basic Energy Sciences, Materials Science and Engineering Division through the Center for Defect Physics, an Energy Frontier Research Center (AA was with partial post-doc support, SNK thesis support, DMN and DDJ are co-PIs), and (SNK post-doc support, DDJ is PI) through Ames Laboratory (DE-AC02-07CH11358 for materials discovery, and DE-FG02-03ER46026 for supplemental code). DOE-funded research was performed at Ames Laboratory, operated for the U.S. Department of Energy by Iowa State University under contract DE-AC02-07CH11358.

\section{Appendix: Derivation of Eq. \ref{eq9}}
To derive Eq. \ref{eq9}, consider the matrix representation of Eq. \ref{eq2}, \ref{eq4}, \ref{eq5} and \ref{eq7}. In the limit $\delta_l \rightarrow 0$, the sine and the cosine matrices {\bf S} and {\bf C} goes to, respectively, zero and unitary  matrix, where $|{\bf C}| \sim |e^{i\phi}| = 1$. Then, accounting for cancellations of sine matrices in the numerator and denominator, we find
\begin{eqnarray}
{\bf Zt} = -{\bf (n S - j C)} e^{+i\delta_l} \xrightarrow[\delta_{l} \rightarrow 0]{} {\bf j}(\kappa r)
\end{eqnarray}
Similarly,
\begin{eqnarray}
{\bf ZtZ} &=& - \kappa \frac{{\bf (n S - j C) (n S - j C)} e^{+i\delta_l}}{\bf S}\nonumber\\
{\bf ZJ} &=& \kappa \frac{{\bf (n S - j C)} {\bf j} }{\bf S}\nonumber
\end{eqnarray}
As $\delta_l \rightarrow 0$, and we apply limit evaluation rules, we find:
\begin{eqnarray}
{\bf ZtZ} &\xrightarrow[\delta_{l} \rightarrow 0]{}& -\kappa\ \left[ i\ {\bf j}(\kappa r)\ {\bf j}(\kappa r) - 2\ {\bf j}(\kappa r)\ {\bf n}(\kappa r) \right]\nonumber\\
{\bf ZJ} &\xrightarrow[\delta_{l} \rightarrow 0]{}&  +\kappa\ {\bf j}(\kappa r)\ {\bf n}(\kappa r)\nonumber
\end{eqnarray}
Therefore,
\begin{eqnarray}
{\bf ZtZ - ZJ} \xrightarrow[\delta_{l} \rightarrow 0]{}  - \kappa\ {\bf j}(\kappa r) \left[ i\ {\bf j}(\kappa r) - {\bf n}(\kappa r) \right] 
\end{eqnarray}


\end{document}